\documentclass[5p,sort&compress]{elsarticle}
\usepackage{amssymb}
\usepackage{amsmath}
\usepackage{graphicx}
\usepackage{hyperref}

\newcommand{\GeV}{\text{\,GeV}}

\newcommand{\e}{\mathrm{e}}

\newcommand{\be}{\begin{equation}}
\newcommand{\ee}{\end{equation}}
\newcommand{\ba}{\begin{eqnarray}}
\newcommand{\ea}{\end{eqnarray}}
\newcommand{\MSb}{$\overline{\text{MS}}$}

\begin{document}

\title{Higgs inflation at the critical point}

\author[cern,uc,bnl]{Fedor Bezrukov}
\ead{Fedor.Bezrukov@uconn.edu}
\author[cern,epfl]{Mikhail Shaposhnikov}
\ead{Mikhail.Shaposhnikov@epfl.ch}

\address[cern]{CERN, CH-1211 Gen\`eve 23, Switzerland}
\address[uc]{Physics Department, University of Connecticut, Storrs, CT
  06269-3046, USA}
\address[bnl]{RIKEN-BNL Research Center, Brookhaven National
  Laboratory, Upton, NY 11973, USA}
\address[epfl]{
Institut de Th\'{e}orie des Ph\'{e}nom\`{e}nes Physiques, 
\'{E}cole Polytechnique F\'{e}d\'{e}rale de Lausanne, CH-1015 Lausanne, 
Switzerland}


\begin{abstract}
  Higgs inflation can occur if the Standard Model (SM) is a
  self-consistent effective field theory up to inflationary scale.
  This leads to a lower bound on the Higgs boson mass, $M_h \geq
  M_{\text{crit}}$. If $M_h$ is more than a few hundreds of MeV above
  the critical value, the Higgs inflation predicts the universal
  values of inflationary indexes, $r\simeq 0.003$ and $n_s\simeq
  0.97$, independently on the Standard Model parameters. We show that
  in the vicinity of the critical point $M_{\text{crit}}$ the
  inflationary indexes acquire an essential dependence on the mass of
  the top quark $m_t$ and $M_h$. In particular, the amplitude of the
  gravitational waves can exceed considerably the universal value.
\end{abstract}

\maketitle

\section{Introduction}

The most economic inflationary scenario is based on the identification
of the inflaton with the SM Higgs boson~\cite{Bezrukov:2007ep} and the
use of the idea of chaotic initial conditions \cite{Linde:1983gd}. The
theory is nothing but the SM with the non-minimal coupling of the
Higgs field to gravity with the gravitational part of the action
\be 
   S_G =\int d^4x \sqrt{-g} \Bigg\{-\frac{M_P^2}{2}R
  - \frac{\xi h^2}{2}R
  \Bigg\}.
  \label{action}
\ee
Here $R$ is the scalar curvature, the first term is the standard
Hilbert-Einstein action, $h$ is the Higgs field, and $\xi$ is a new
coupling constant, fixing the strength of the ``non-minimal''
interaction. The  presence of non-minimal coupling is  required for
consistency of the SM in curved space-time (see,
e.g.~\cite{Feynman:1996kb}). The value of $\xi$ cannot be fixed
theoretically within the SM.

The presence of the non-minimal coupling insures the flatness of the 
scalar potential in the Einstein frame at large values of the Higgs
field. If radiative corrections are ignored, the successful inflation
occurs for any values of the SM parameters provided  $\xi \simeq 
47000\sqrt{\lambda}$, where $\lambda$ is the Higgs boson
self-coupling. This condition comes from the requirement to have the
amplitude of the scalar perturbations measured by the COBE satellite.
After fixing the unknown constant $\xi$ the theory is completely
determined. It predicts the tilt of the scalar perturbations  given by
$n_s\simeq 0.97$ and the tensor-to-scalar ratio $r\simeq 0.003$. After
inflationary period, the Higgs field oscillates, creates particles of
the SM, and produces the Hot Big-Bang with initial temperature in the
region of $10^{13\text{--}14}$\GeV{}
\cite{Bezrukov:2008ut,GarciaBellido:2008ab}.

The quantum radiative corrections can change the form of the effective
potential and thus modify the predictions of the Higgs inflation. The
most significant conclusion coming from the analysis of the quantum
effects is that the Higgs inflation can only take place if the mass of
the Higgs boson is greater than some critical number
$M_{\text{crit}}$ 
\cite{Bezrukov:2008ej,DeSimone:2008ei,Barvinsky:2008ia,
Bezrukov:2009db,Barvinsky:2009fy},
\begin{align}
  M_h>M_{\text{crit}}.
  \label{inflcond}
\end{align}
Roughly speaking, the Higgs
self-coupling constant must be positive at the energies up to the
inflationary scale, leading to this constraint. In numbers 
\cite{Bezrukov:2012sa,Degrassi:2012ry,Buttazzo:2013uya},
\begin{align}
\nonumber
M_{\text{crit}}= \Big[129.6 + \frac{y_t^\text{phys} - 
0.9361}{0.0058}\times 2.0 -\\
\frac{\alpha_s-0.1184}{0.0007}\times
0.5\Big]\GeV. 
\label{mcrit}
\end{align}
Here $y_t^\text{phys}$ is the top Yukawa coupling in \MSb{}
renormalisation scheme taken at $\mu_t=173.2\GeV$\footnote{For precise
  relation between $y_t^\text{phys}$ and the pole top mass $m_t$ see
  \cite{Bezrukov:2012sa,Buttazzo:2013uya} and references therein.},
$y_t^\text{phys}\equiv y_t^\text{phys}(\mu_t)$ and $\alpha_s$ is the
QCD coupling at the $Z$-boson mass.  Thanks to complete two-loop
computations of \cite{Buttazzo:2013uya} and three-loop beta functions
for the SM couplings found in
\cite{Mihaila:2012fm,Mihaila:2012pz,Chetyrkin:2012rz,Chetyrkin:2013wya,
  Bednyakov:2012en,Bednyakov:2013eba} this formula may have a very
small theoretical error, $0.07$\GeV, with the latter number coming
from an ``educated guess'' estimates of even higher order terms (see
the discussion in \cite{Bezrukov:2012sa} and more recently in
\cite{Shaposhnikov:2013ira}). The main uncertainty in determination of
$M_{\text{crit}}$ is associated with experimental and theoretical
errors in determination of $y_t^\text{phys}$ from available data.
Accounting for those, the value of $M_{\text{crit}}$ is about 2
standard deviations from the mass of the Higgs boson observed
experimentally at CERN \cite{Aad:2012tfa,Chatrchyan:2012ufa}.

The determination of the inflationary indexes accounting for radiative
corrections is somewhat more subtle and depends on the way the quantum
computations are done (the SM with gravity is non-renormalizable, what
introduces the uncertainty). In \cite{Bezrukov:2008ej,Bezrukov:2009db}
we formulated the natural subtraction procedure (called ``prescription
I'') which uses the field independent subtraction point in the
Einstein frame (leading to scale-invariant quantum theory in the
Jordan frame for large Higgs backgrounds) and computed $n_s$ and $r$
for the Higgs masses that exceeded $M_{\text{crit}}$ by just a small
amount of few hundreds of MeV\footnote{We also performed the
  computation with the use of another subtraction procedure (called
  ``prescription II''), which has a field-independent subtraction
  point in the Jordan frame
  \cite{Barvinsky:2008ia,DeSimone:2008ei,Barvinsky:2009fy}.}. We have
shown that the values of $n_s$ and $r$ are remarkably stable in this
domain and coincide with the tree estimates. However, we did not
analyse what happens in the close vicinity of the critical point.
Partially, this has been studied in \cite{Allison:2013uaa}, but the
peculiar inflationary behaviour found in the present work was not
discussed in \cite{Allison:2013uaa}.

The aim of the present paper is to study the behaviour of the
inflationary indexes close to the critical point. In what follows we
will use the prescription I. We expect to have qualitatively the same
results in the prescription II, though the numerical values will be
somewhat different. We will see that $n_s$ and $r$ acquire a strong
dependence on the mass of the Higgs boson and the mass of the top
quark. Thus, if the cosmological observations will show that one or
both indexes do not coincide with those given by the tree analysis,
they will indicate that in instead of inequality (\ref{inflcond}) we
must have an equality between the Higgs mass and its critical value,
$M_h=M_{\text{crit}}$.

\section{The critical point}

The behaviour of the scalar self-coupling constant $\lambda$ as a
function of the \MSb{} parameter $\mu$ (energy)  in the SM is very
peculiar. If the mass of the top quark and of the Higgs boson are
varied within their experimentally allowed intervals, 
it can be
approximated in the region of Planck energies
($M_P=2.44\times10^{18}$\GeV) with a good accuracy as follows:
\be
\lambda(z)=\lambda_0 + b \left(\log z \right)^2,
\ee
where 
\be
z=\frac{\mu}{qM_P},
\ee
$\lambda_0$, $q$ and $b$ are some functions of the top quark (pole)
mass, Higgs mass, and the strong coupling constant $\alpha_s$, see
Section \ref{conn}.  It happens that $\lambda_0$ is small,
$\lambda_0\ll 1$ and $q$ is of the order of one\footnote{A possible
  explanation of these facts may lie in the asymptotic safety of the
  SM \cite{Shaposhnikov:2009pv}.  This is also a basis of the
  ``multiple point principle'' of \cite{Froggatt:1995rt}.}. To put it
in words, both the value of $\lambda$ and of its beta-function,
$\beta_\lambda = \mu \partial \lambda/\partial \mu$, are close to zero
near the Planck scale. It is this fact that changes the behaviour of
the inflationary indexes, as is demonstrated below.

The renormalisation group improved effective potential in the Einstein
frame with an accuracy sufficient for the present discussion can be
written as follows \cite{Bezrukov:2009db}:
\begin{equation}
  \label{U(chi)}
  U(\chi) \simeq \frac{\lambda(z') }{4\xi^2}\bar\mu^4.
\end{equation}
Here
\be
z'=\frac{\bar\mu}{\kappa M_P}
\ee
and
\be
\bar\mu^2=M_P^2 \left(1-\e^{-\frac{2\chi}{\sqrt{6}M_P}} \right),
\ee
where $\chi$ is the canonically normalised scalar field related to the
original Higgs field by a known transformation \cite{Bezrukov:2007ep}
(see eq. (\ref{deriv})).  The \MSb{} parameter $\mu$ that optimises
the convergence of the perturbation theory is related to $\bar\mu$ as
\begin{equation}
  \label{renormcondition}
  \mu^2 =   \alpha^2\frac{y_t(\mu)^2}{2}\frac{\bar\mu^2 }{\xi(\mu)}
\end{equation}
with $\alpha \simeq 0.6$.  This numerical value follows from the
minimisation of the one-loop Coleman-Weinberg effective potential
\cite{Bezrukov:2008ej,Bezrukov:2009db} (two-loop contributions do not
further change $\alpha$ significantly).  The expression (\ref{U(chi)})
is valid for $\xi > 1$, and the scale dependence of $\xi$ can be
neglected.
  
\begin{figure}[!htb]
\centering
\includegraphics*[width=\columnwidth]{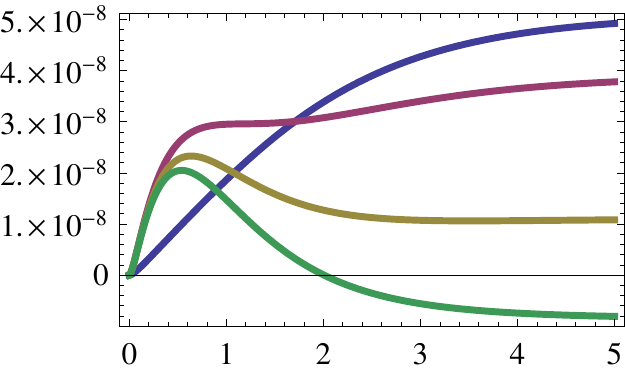}
\caption{The schematic change of the form of the effective potential
depending on $\lambda_0$. For better visibility the values of $\xi$
are different for different lines. The horisontal axis corresponds to
the canonically normalized field $\chi$, the vertical axis to the
effective potential, all in Planck units.}
\label{fig:pot}
\end{figure}

Let us now consider the change of the form of the potential if
$\lambda_0$, $\kappa$ and $\xi$ are varying. For $\lambda_0 \gg b/16$ the
potential is a rising function of the field $\chi$, realizing the
``tree'' Higgs inflation (see Fig.~\ref{fig:pot}, blue curve). If
$\lambda_0 = b/16$, a new feature appears: the first and the second
derivatives of the potential potential are equal to zero at some point
(see Fig.~\ref{fig:pot}, red curve).  For $\lambda_0 < b/16$ but still
close to $b/16$ we get a wiggle on the potential, which is converted
into a maximum for somewhat smaller $\lambda_0$ (see
Fig.~\ref{fig:pot}, brown line).  Decreasing $\lambda_0$ even further
leads to the unstable electroweak vacuum, Fig.~\ref{fig:pot} (green
line).  Clearly, the necessary condition for inflation to happen in
the slow-roll regime is to have $dV(\chi)/d\chi >0$ for all $\chi$,
i.e.\ the absence of a wiggle. For $\lambda_0 \gg b/16$ all the
potentials are very much similar, leading to the independence of
inflationary indexes on the parameters, while if $\lambda_0$ is close
to $b/16$, the form of the potential changes, and the dependence of
$r$ and $n_s$ on $\lambda_0$ and $\kappa$ (and, therefore on $M_h$ and
$m_t$) shows up.

The parameter $\kappa$ controls the value of $\chi$ where the wiggle
would appear for $\lambda_0 = b/16$, the parameter $\lambda_0-b/16$
tells how close we are to the appearance of the feature, while a
combination of $\lambda_0$ and $\xi$ determines the asymptotic of the
potential at large $\chi$.  Let us note that inflation with large $r$
in the potentials with near vanishing $U'$ at some value of the field
along the inflationary slow-roll evolution was considered in
\cite{BenDayan:2009kv,Hotchkiss:2011gz}.

\section{The inflationary indexes}

Once the potential is known it is straightforward to determine
inflationary indexes. For this end it is more convenient to use the
three parameters $\kappa$, $\lambda_0$ and $\xi$ as independent
variables, characterising the physics at the inflation scale (the
relation to low energy observables will be discussed in Section
\ref{conn}).  Note that the value of $b$ is stable against the change
of the SM parameters in the vicinity of the critical point and can be
fixed as $b\simeq 2.3\times 10^{-5}$.

\begin{figure}[!htb]
  \centering
  \includegraphics*[width=\columnwidth]{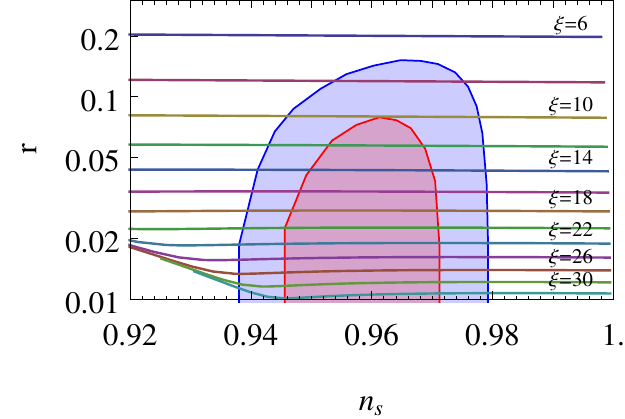}
  \caption{The dependence of the inflationary indexes $n_s$ and $r$ on
    $\xi$ and $\kappa$, the parameter $\lambda_0$ is fixed by the COBE
    normalisation. Along the nearly horisontal lines $\xi$ is fixed
    and $\kappa$ is varying within the interval $\{0.9,1.1\}$.  We
    also show 1 and 2 $\sigma$ contours coming from the results of
    Planck \cite{Ade:2013uln}.}
  \label{fig:xigrid}
\end{figure}

\begin{figure}[!htb]
  \centering
  \includegraphics*[width=\columnwidth]{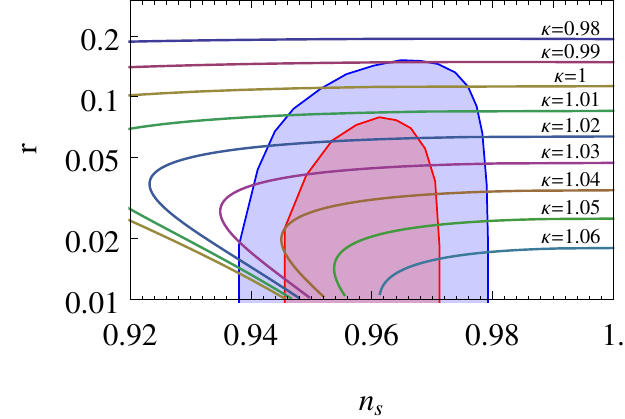}
  \caption{The same as in Fig.~\ref{fig:xigrid}, but with the grid of
    constant $\kappa$ lines.  The parameter $\xi$ is varying within
    the interval $\{5,30\}$. }
  \label{fig:kappagrid}
\end{figure}

In the Higgs inflation far from the critical point the parameter
$\kappa$ is irrelevant, whereas $\lambda_0$ and $\xi$ always appear in
the combination $\lambda_0/\xi^2$, meaning that the potential depends
on one parameter only. Fixing it from the COBE normalisation then
leads to prediction of $n_s$ and $r$. Close to the critical point the
situation is changed - all the three parameters are now essential. The
parameter counting leads to the conclusion that any values of $n_s$
and $r$ are now possible. This expectation is confirmed by a detailed
analysis, see Figs.~\ref{fig:xigrid} and \ref{fig:kappagrid}.

In Fig.~\ref{fig:xigrid} (\ref{fig:kappagrid}) we show the lines of
constant $\xi$ ($\kappa$) on the plane $(n_s,r)$, the parameter along
the line is associated with the variation of $\kappa$ ($\xi$) within
the interval $\{0.9,1.1\}$ ($\{5,30\}$).


\begin{figure}[!htb]
  \centering
  \includegraphics*[width=\columnwidth]{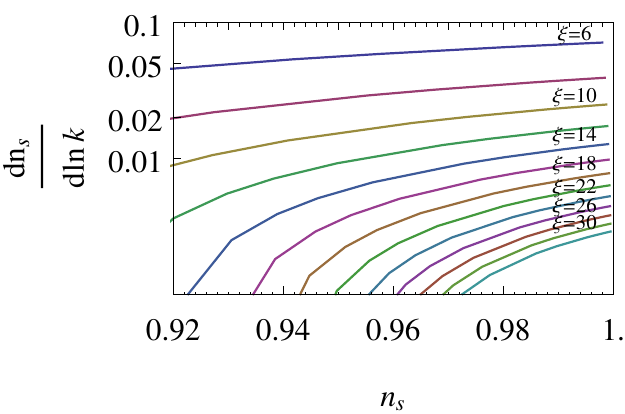}
  \caption{The same as in Fig.~\ref{fig:xigrid}, but now on the plane
    $(n_s, dn_s/d\ln k)$, which includes the running of the scalar
    spectral index.}
  \label{fig:runninggrid}
\end{figure}

In Fig.~\ref{fig:runninggrid} we show the running of the scalar index
as a function of $\kappa$ and $\xi$. One can see that it is positive.
A word of caution should be said in this place, warning from the use
of the available inflationary constraints from CMB measurement.  With
the potential (\ref{U(chi)}) close to the critical point the behaviour
of the spectral index in the observable inflationary region is
complicated (i.e.\ expansion in terms of running and running of the
running of the spectral index over the observable inflationary window
is not a good approximation, contrary to the case of power law
potentials).  Thus it is necessary to make the complete fit of the
perturbation spectrum, generated form the potential (\ref{U(chi)}) to
the CMB observations, like it is suggested in
\cite{Lesgourgues:2007gp,Lesgourgues:2011re,Blas:2011rf}.  This
analysis goes beynod the scope of the present letter.

\begin{figure}[!htb]
  \centering
  \includegraphics*[width=\columnwidth]{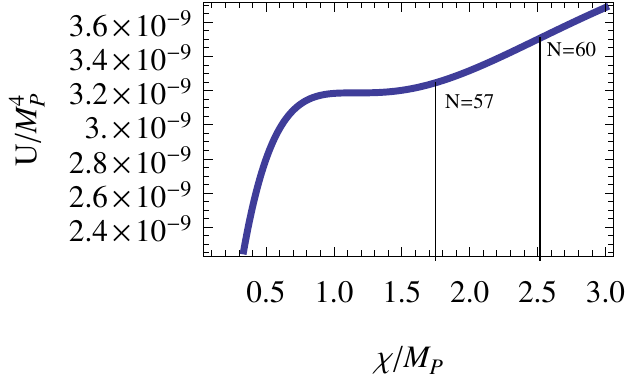}
  \caption{The form of the effective potential which leads to $r=0.1$,
    $n_s=0.96$.  The field values corresponding to the $N=57$ and $N=60$
    e-foldings are marked by vertical lines, roughly indicating the
    observable window for inflation.}
  \label{fig:bicep}
\end{figure}

The picture of the Higgs-inflation potential which gives
$r=0.1$, $n_s=0.96$ is shown in Fig.~\ref{fig:bicep}. 

A very interesting feature of the inflation near the critical point is
the drastic decrease of the necessary non-minimal coupling $\xi$ down
to a number of the order of ten. The large value of $\xi$, necessary
for the Higgs inflation far from the critical point, effectively
introduces a new strong-coupling threshold $\Lambda \sim M_P/\xi$ well
below the Planck scale, if the scattering of the SM particles is
considered around the EW vacuum \cite{Burgess:2009ea,Barbon:2009ya}.
Though this fact does not invalidate the self-consistency of the Higgs
inflation \cite{Bezrukov:2010jz, Ferrara:2010in} which occurs at large
Higgs fields, it requires the UV completion of the SM or self-healing
of high energy scattering \cite{Aydemir:2012nz,Calmet:2013hia} at energies much
smaller than the Planck scale. The Higgs inflation at the critical
point does not require any new cutoff scale, different from the Planck
scale.

The evolution of the Universe after the Higgs inflation at the
critical point is different from that for the case $\xi \gg 1$.  If
$\xi \gg 1$, the Universe after inflation is ``matter dominated'' due
to oscillations of the Higgs field. The transition to the radiation
dominated Universe occurs due to particle production after some time,
but not later than  after  ${\cal O}(\xi/2\pi)$ oscillations
\cite{Bezrukov:2008ut,GarciaBellido:2008ab}. For $\xi \sim 10$ we have
the radiation-dominated epoch right after inflation is finished. Of course, the 
system will come to thermal equilibrium  only after a number of oscillations 
of the Higgs field.

\section{Connection between low energy and high energy observables}
\label{conn}

The aim of this section is elucidating the relation between the
parameters of the inflationary potential $\kappa$ and $\lambda_0$ and
top quark and Higgs masses measured in low energy experiments. The
main problem here is that the Standard Model in the Einstein frame is
essentially non-polynomial and thus non-renormalizable, meaning that
the required connection cannot be found without extra assumptions
about the structure of the underlying fundamental theory
\cite{Bezrukov:2010jz}.

The most conservative (and thus most predictive ) hypothesis of the
absence of new physics between the Planck and Fermi scales still has
the uncertainties in the relation between low energy and high energy
parameters \cite{Bezrukov:2010jz}.  To discuss these uncertainties let
us consider (the most important numerically) interactions of the top
quark with the Higgs field, given in the Einstein frame by
\be
L_t = \frac{y_t}{\sqrt{2}} \bar t t F(\chi).
\ee
Here 
\be
F(\chi)= \frac{h}{\Omega}~,
\ee
$\Omega^2=1+\xi h^2/M_P^2$ is the conformal factor,
and the canonically normalised field $\chi$ is related to the Higgs field $h$ via
\be
\frac{dh}{d\chi}=\frac{\Omega^2}{\sqrt{\Omega^2 + \xi(6\xi+1) h^2/M_P^2}}.
\label{deriv}
\ee
To remove the divergencies in the arbitrary background fields, the
counter-terms must be added to the action. The scalar correction to
the $\bar t t h$ vertex requires the modification of $y_t$ as
\be
y_t  \to y_t + \frac{y_t^3}{16\pi^2}\left(\frac{3}{\epsilon} + C_t \right) F'^2
\ee
and the top quark loop contribution to scalar self-interaction gives
\be
\lambda  \to \lambda - \frac{y_t^4}{16\pi^2}\left(\frac{6}{\epsilon} - C_\lambda \right) F'^4,
\ee
where $\epsilon$ is the parameter of dimensional regularisation, $C_t$
and $C_\lambda$ are the (arbitrary) constant parts of the
counter-terms, and $F'=dF/d\chi$.  These constant parts cannot be
fixed theoretically with the use of the SM Lagrangian and must be
found from observations or from (unknown yet) UV complete theory hosting the SM at
low energies.

For the small Higgs backgrounds $h \ll M_P/\xi$ the derivative $F'$ is
close to one, and the parameters $C_t$ and $C_\lambda$ are absorbed in
the definition of low-energy top Yukawa coupling and scalar
self-coupling and thus are unobservable at low energies.  However, in
the inflationary region, at $h > M_P/\xi$, the derivative $F'$ goes to
zero, and it is the couplings $y_t$ and $\lambda$, rather than
$y_t^\text{phys}=y_t+ \frac{y_t^3}{16\pi^2} C_t $ and
$\lambda^\text{phys}= \lambda - \frac{y_t^4}{16\pi^2} C_\lambda$, that
contribute to cosmological observables. The transition between two
regimes (for $\xi>1$) occurs approximately at $h^*=
\frac{1}{2\sqrt{6}}\frac{M_P}{\xi}$, corresponding to the fastest
falloff of the function $F'^2$.  It can be well approximated by a
sudden jump of the coupling constant from $y_t^\text{phys}$ to $y_t$
at $h=h^*$. The obvious inflationary requirement is that in the domain
$h<h^*$ the {\em physical, low energy} $\lambda^\text{phys}$ must be
positive, i.e.\ the inequality (\ref{inflcond}) must be valid.

To determine the parameters of the inflationary potential
(\ref{U(chi)}) we define the ``inflationary'' top and Higgs masses
$m_t^*$ and $M_h^*$ that lead to $y_t$ and $\lambda$ at high energies
through the renormalisation group evolution (SM running up to $h^*$
and chiral SM running afterwords \cite{Bezrukov:2009db}),
\emph{without any jumps} at $h=h^*$. The inflationary masses are
related to the physical masses $m_t$ and $M_h$ as follows:
\be
m_t^* = m_t \left(1- \frac{y_t^2C_t}{16\pi^2} \right)
,\quad
M_h^*=M_h\left(1-  \frac{y_t^4C_\lambda }{16\pi^2} \frac{h_0^2}{M_h^2}\right),
\ee
where $h_0=250$ GeV is the vacuum expectation value of the Higgs
field, and all the constants are taken at low energy scale.  The
presence of the unknown coefficients $C_t$ and $C_\lambda$ results
exactly from the unremovable uncertainty following from the
non-renormalizable character of the SM coupled to gravity in the
non-minimal way. Numerically (in the units of GeV, and for $M_h\simeq
126$\GeV),
\be
m_t^* \simeq  m_t - C_t,~~M_h^* \simeq  M_h - 3 C_\lambda.
\ee

Now, we can combine the discussion of inflation with the low energy
parameters accounting for the uncertainties discussed above.  The
functions $\lambda_0$, $q$ and $b$ can be expressed through the high
energy inflationary parameter and can be found from the analysis of
the renormalisation group running for $\lambda$. The fitting formulas
are given below\footnote{We used the analysis made in
  \cite{Bezrukov:2012sa} to produce the fitting formulas and fixed
  $\alpha_s=0.1184$. The account of more precise mapping at the
  electroweak scale made in \cite{Buttazzo:2013uya} can change the
  extraction of the Higgs and top masses from the cosmological data by
  amount of ${\cal O}(100)$ MeV.}:
\begin{align}
  \lambda_0 = &\ 0.003297 \big((M_h^*-126.13) 
               - 2 (m_t ^*-171.5) \big), \notag\\
  \nonumber
  q =         &\ 0.3 \exp \big(0.5(M_h^*-126.13)-0.03(m_t^*-171.5) \big),
\\
 \nonumber
  b = &\ 0.00002292 
       - 1.12524\times10^{-6}\big( (M_h^*-126.13) \\
 &-
 1.75912 (m_t^*-171.5)\big), 
\end{align}
where  $M_h^*$ and $m_t^*$ are to be taken in GeV.

\begin{figure}[!htb]
\centering
\includegraphics[width=\columnwidth]{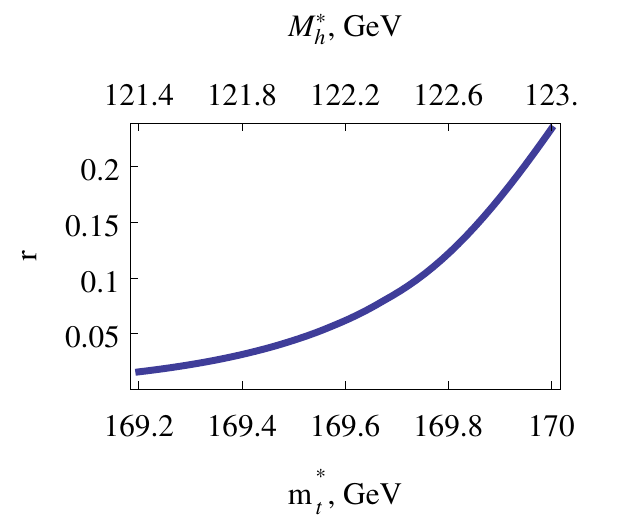}
\caption{The dependence of the tensor-to-scalar ratio $r$ on the high
  energy inflationary Higgs boson and top quark masses $M_h^*$ and
  $m_t^*$. The value of $n_s$ along the curve is vithin the interval
  $n_s\in\{0.94,0.98\}$.}
\label{fig:rns}
\end{figure}

\begin{figure}[!htb]
\centering
\includegraphics[width=\columnwidth]{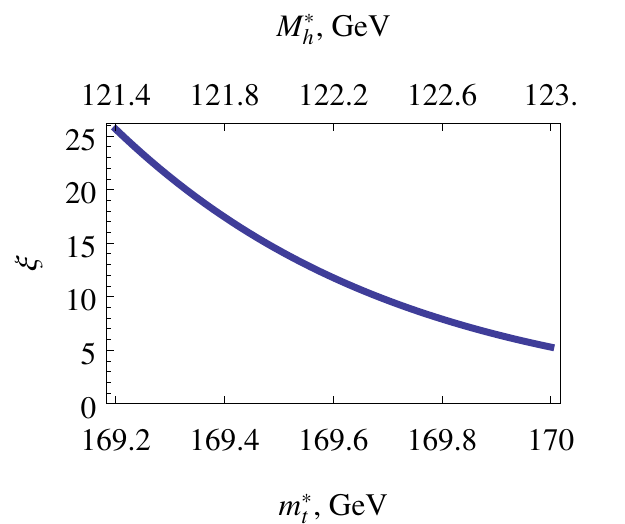}
\caption{The dependence of the required non-minimal coupling on the
high energy inflationary Higgs boson and top quark masses $M_h^*$ and $m_t^*$.}
\label{fig:xi}
\end{figure}

\begin{figure}[!htb]
\centering
\includegraphics*[width=\columnwidth]{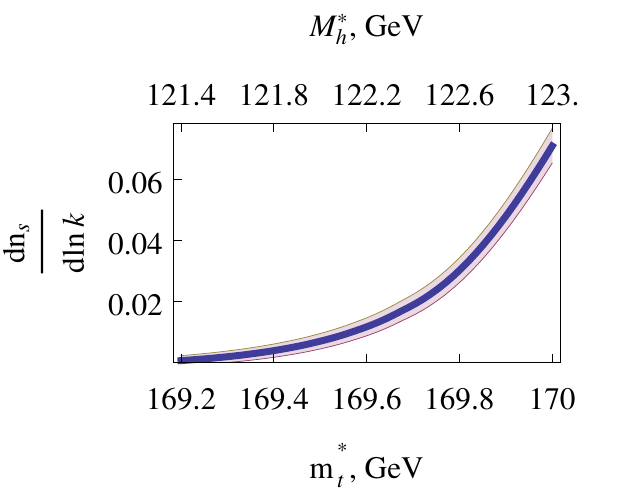}
\caption{The same as in Fig.~\ref{fig:rns}, but for the running of the
  scalar spectral index.  The shaded area corresponds to change of
  $\kappa$ leading to change of $n_s$ within the Planck allowed values
  (for Figs.~\ref{fig:rns} and \ref{fig:xi} this area is within the
  thickness of the lines on the plots).}
\label{fig:running}
\end{figure}

These equations can be used now to determine the dependence of the
cosmological parameters on $m_t^*$ and $M_h^*$. Since the value of the
scalar tilt $n_s$ depends very strongly on $\kappa$, we fix it in the
experimentally allowed region $n_s\in\{0.94,0.98\}$ and present in
Figs.~\ref{fig:rns} and \ref{fig:xi} the dependence of $r$ and
required $\xi$ on high energy values of the top quark and Higgs
masses. Also, Fig.~\ref{fig:running} shows the running of the spectral
index.

The combination of particle physics and cosmological measurements
allows to fix unknown parameters $C_t$ and $C_\lambda$. If we take,
for instance, $r=0.12$, then $M_h^* \simeq 122.6$ GeV and $m_t^*\simeq
169.8$ GeV, $\xi \simeq 8$. To get the physics low energy value of the
Higgs mass $M_h= 125.6$ GeV we need $C_\lambda\simeq 1$. The value of
$C_t \simeq 1.5$ would bring the physical top mass to $m_t \simeq
171.5$ GeV, consistent with the measured top quark mass within 2
$\sigma$ uncertainties. The relative jump of the top Yukawa coupling
at $h=h^*$ is $(y_t^\text{phys}-y_t)/y_t \simeq 0.024$, while the jump
in $\lambda$ is very sensitive to the physical Higgs and top masses
and can be made equal to zero by tuning $m_t$ and $M_h$ within few
MeV.

\section{Conclusions}

The Higgs inflation for $M_h>M_{\text{crit}}$ is a predictive theory
for \emph{cosmology}, as the values of the inflationary indexes are
practically independent of the SM parameters. Near the critical point
the situation completely changes, and we get a strong dependence of
$n_s$ and $r$ on the precise values of the \emph{inflationary masses}
of the top quark and the Higgs boson $m_t^*$ and $M_h^*$.  In this
regime the Higgs inflation becomes a predictive theory for \emph{high
  energy domain of particle physics}, as any deviation of inflationary
indexes from the tree values tells that we are at the critical point,
fixing thus the inflationary values of masses of the top quark and the
Higgs boson $m_t^*$ and $M_h^*$.  It is amazing that a possible
detection of large tensor-to-scalar ratio $r$ in \cite{Ade:2014xna}
gives the inflationary top quark and Higgs boson masses close to their
experimental values $m_t$ and $M_h$. This tells that the uncertainties
related to the transition from low and high energies corresponding to
the Higgs field $h^* \propto M_P/\xi$ are quite small.

We conclude with a word of caution. All results here are based on the
assumption of the validity of the SM up to the Planck scale. If this
hypothesis is removed, the Higgs inflation remains a valid
cosmological theory, but its predictability is lost even far from the
critical point. For example, the modification of the kinetic term of
the Higgs field at large values of $H$, leads to a considerable
modification of $r$
\cite{Germani:2010gm,Germani:2010ux,Nakayama:2014koa} (see also
\cite{Kamada:2010qe,Kamada:2012se,Kamada:2013bia} for generalized
Higgs inflation with Horndenski type terms).  The change of the
structure of the Higgs-gravity interaction to, for instance,
\be
M_P^2 R \sqrt{1+\xi |H|^2/M_P^2}, 
\ee
will make the potential in the Einstein frame quadratic with respect
to the field $\chi$ and thus would modify $r$ and $n_s$, making them
the same as in the chaotic inflation with free massive scalar field.
Another assumption is about the absense of operators suppressed by the
Planck scale (or various tree level unitarity violation scales
\cite{Bezrukov:2010jz}), which may be justified by a special scale (or
shif in the Einstein frame) symmetry of the gravitational physics.
Adding them would change the inflationary physics,
cf.\ \cite{Branchina:2013jra} for importance of such terms for the
stability of electroweak vacuum.

While this paper was in preparation, the article \cite{Hamada:2014iga}
appeared, where the possibility to have large value of $r$ for the
Higgs inflation close to the critical point was also pointed out.

\bigskip

The authors would like to thank CERN Theory Division, where this paper
was written, for hospitality.  We thank Dmitry Gorbunov for helpful
comments. The work of M.S. is supported in part by the European
Commission under the ERC Advanced Grant BSMOXFORD 228169 and by the
Swiss National Science Foundation.

\bibliography{Papers}
\bibliographystyle{elsarticle-num}

\end{document}